\documentclass[10pt]{amsart}

\newtheorem{theorem}{Theorem}[section]
\newtheorem{lemma}[theorem]{Lemma}
\newtheorem{corollary}[theorem]{Corollary}
\newtheorem{proposition}[theorem]{Proposition}

\begin{document}

\title{Codekets}

\author{Mihai Caragiu}
\address{\newline Mihai Caragiu \newline Department of Mathematics
\newline Ohio Northern University, Ada, OH 45810 \newline {\rm \it E-mail addres: m-caragiu1@onu.edu}}
\curraddr{}
\thanks{}

\date{}

\dedicatory{}

\begin{abstract}
To every binary linear $[n,k]$ code $C$ we associate a quantum
state $|\Psi _C\rangle \in H^{\otimes n}$, where $H$ is the
two-dimensional complex Hilbert space associated to the spin
$\frac {1}{2}$ particle. We completely characterize the
expectation values of the products of $x$-, $y$- or $z$- spins
measured in the state $|\Psi _C\rangle $, for each of the
particles in a chosen subset. This establishes an interesting
relationship with the dual code $C^{\perp }$. We also address the
case of nonlinear codes, and derive both a bound satisfied by the
expectations of spin products as well as a nice algebraic
identity.

\end{abstract}

\maketitle

\bigskip

\tableofcontents

\newpage

\

\section { Codekets}

\

In the present paper we will prove a duality property holding for
a class of entangled systems of spin $\frac {1}{2}$ fermions. For
a single such particle, the Hilbert space $H$ associated to the
spin has an orthonormal basis consisting of eigenvectors of
$\sigma _z$, namely
$|+\rangle = \left(%
\begin{array}{c}
 1 \\
 0 \\
\end{array}%
\right)\ \ {\rm and} \ \ |-\rangle = \left(%
\begin{array}{c}
 0 \\
 1 \\
\end{array}%
\right)$. In order to set up a consistent binary notation we will
use $|0\rangle $ for $|+\rangle $ and $|1\rangle $ for $
|-\rangle$, $\sigma _{0,0}$ for ${\bf 1}_H$ , $\sigma _{0,1}$ for
$\sigma _x$, $\sigma _{1,1}$ for $\sigma _y$, and $\sigma _{1,0}$
for $\sigma _z$, where ${\bf 1}_H$ is the identity transformation
of $H$, while $\sigma_x, \sigma _y,\sigma _z$ are the Pauli
matrices (see [2], Chapter 4). Thus,

$$
\sigma _{0,0}=\left(%
\begin{array}{cc}
 1 & 0 \\
 0 & 1 \\
\end{array}%
\right)\ \ , \ \
\sigma _{0,1}=\left(%
\begin{array}{cc}
 0 & 1 \\
 1 & 0 \\
\end{array}%
\right)\ \ , \ \
\sigma _{1,1}=\left(%
\begin{array}{cc}
 0 & -i \\
 i & 0 \\
\end{array}%
\right)\ \ , \ \
\sigma _{1,0}=\left(%
\begin{array}{cc}
 1 & 0 \\
 0 & -1 \\
\end{array}%
\right).$$

\

With the binary notation we just introduced, the following
relations hold:

$$\sigma_{0,0}|0\rangle =|0 \rangle \ \ {\rm and} \ \ \sigma_{0,0}|1\rangle =|1\rangle ,\leqno (1)$$
$$\sigma_{0,1}|0\rangle =|1\rangle \ \ {\rm and } \ \ \sigma_{0,1}|1\rangle =|0\rangle ,\leqno (2)$$
$$\sigma_{1,1}|0\rangle =i|1\rangle \ \ {\rm and} \ \ \sigma_{1,1}|1\rangle =-i|0\rangle .\leqno (3)$$
$$\sigma_{1,0}|0\rangle =|0 \rangle \ \ {\rm and} \ \ \sigma_{1,0}|1\rangle =-|1\rangle ,\leqno (4)$$

\

\noindent Let $C$ be a linear binary $[n,k]$ code (that is, a
$k$-dimensional subspace of the $n$-dimensional vector space $\{
0,1\}^n$ over the field with $2$ elements). In [1] we associated
to $C$ the following quantum state (element of $H^{\otimes n}$)
describing a system of $n$ spin $\frac {1}{2}$ particles:
$$|\Psi_C\rangle :=\frac {1}{2^{k/2}}\sum_{(x_1,...,x_n)\in C}|x_1\rangle \otimes |x_2\rangle \otimes ... \otimes |x_n\rangle .\leqno (5)$$
Let us agree to call $|\Psi_C\rangle $ the {\it codeket}
associated to the binary code $C$. Note that in the special case
of the binary $[n,1]$ code
$$C=\{ (0,0,...,0),(1,1,...,1)\},$$
the codeket $|\Psi _C\rangle $ is the well known
Greenberger-Horne-Zeilinger ($GHZ$) quantum entangled state [4],
[3]
$$|\Psi_{GHZ}\rangle =\frac {1}{\sqrt 2}\left ( |00...0\rangle + |11...1\rangle \right
).$$

\

\section {An imaginary experiment setup}

\

Assume that for a system in the state defined by the codeket
$|\Psi _C\rangle $ we perform the following imaginary experiment:
we measure either the $x$-component, the $y$-component or the
$z$-component of the spin for each particle in a certain {\it
subset} of the set of $n$ fermions, then we take the product of
the measured values. This product will be either 1 or $-1$. At a
formal level, a particular measurement of $x$-, $y$- or $z$- spins
for a subset of the set of particles in the state (5) can be
represented as a product operator of the form
$$\sigma _{A,B} = \sigma _{a_1,b_1}\otimes \sigma _{a_2,b_2}\otimes ... \otimes \sigma _{a_n,b_n},$$
where $A:=(a_1,...,a_n)\in \{ 0,1\}^n$ and $B:=(b_1,...,b_n)\in \{
0,1\}^n$ codify the measurement choice we make for every particle.
For example, let us assume $n=4$, $A=(1,0,1,0)$ and $B=(0,1,1,0)$.
Then $(a_1,b_1)=(1,0)$, $(a_2,b_2)=(0,1)$, $(a_3,b_3)=(1,1)$ and
$(a_4,b_4)=(0,0)$, so that we measure the $z$-component of the
spin for the first particle, the $x$-component for the second, the
$y$-component for the the third, while there is no measurement
performed on the fourth particle. Thus, every such pair $(A,B)$ of
binary vectors codifies both a subset of the set of $n$ fermions
(in the previous example, the subset consists of the first three
particles), as well as a selection of $x$-, $y$- and $z$- spins to
be measured for each of the particles in that subset. For every
$i$ between $1$ and $n$ we perform a ($x$-, $y$- or $z$-) spin
measurement on the $i$-th particle if and only if $(a_i,b_i)\neq
(0,0)$.

\

\

\centerline {\begin{tabular}{|c|c|}
  \hline
  $(a_i,b_i)$ & Measurement on the $i$-th particle \\
  \hline
  $(0,0)$ & None \\
  $(0,1)$ & $\sigma_x$ \\
  $(1,1)$ & $\sigma_y$ \\
  $(1,0)$ & $\sigma_z$ \\
  \hline
\end{tabular}}

\

\

Note that in [1] we considered the special case in which
measurements of $x$- or $y$- components of the spin were made for
all particles in the system described by the codeket
$|\Psi_C\rangle $. The following lemma provides a unified
description of the action of the operators $\sigma _{a,b}$ on the
vectors $|0\rangle $ and $|1\rangle $.

\begin {lemma} If $a,b,x$ are bits, then
$$\sigma _{a,b}|x\rangle = e^{i\left (\frac {\pi ab}{2}+\pi ax\right )}
|x+b\rangle ,\leqno (6)$$ where the bit sum $x+b$ is taken modulo
2.
\end {lemma}

\begin {proof}
Straightforward verification, based on the relations (1) through
(4).
\end {proof}

\

\section {Expectation values for products of spins}

\

\noindent We will now determine $\Pi (C,A,B)= \langle \Psi _C |
\sigma _{A,B} |\Psi _C \rangle$, the expectation value in the
quantum state defined by the codeket $|\Psi _C\rangle $ of the
product of spins produced by the measurements codified as shown
above by a given pair of binary vectors $(A,B)$. For the relevant
quantum mechanical formalism, see [2], Chapter 2. By using (6) we
calculate the action of $\sigma _{A,B}$ on the codeket
$|\Psi_C\rangle $:

$$\sigma _{A,B} |\Psi _C \rangle = \frac {1}{2^{k/2}}\sum_{X=(x_1,...,x_n)\in
C}|\sigma _{a_1,b_1} x_1\rangle \otimes ... \otimes |\sigma
_{a_n,b_n} x_n\rangle = $$ $$ \frac {1}{2^{k/2}} \sum _{X\in
C}e^{\frac {i\pi }{2}(a_1b_1+...+a_nb_n)+i\pi
(a_1x_1+...+a_nx_n)}|x_1+b_1\rangle \otimes ... \otimes |x_n+b_n
\rangle. $$

\

That is,

$$\sigma _{A,B} |\Psi _C \rangle = \frac {e^{\frac {i\pi }{2} (A|B)}}{2^{k/2}}
\sum_{X\in C}e^{i\pi (A|X)}|X+B\rangle , \leqno (7)$$ where
$(A|X)=a_1x_1+...+a_nx_n$ and $|X+B\rangle =|x_1+b_1\rangle
\otimes ... \otimes |x_n+b_n\rangle$, where each $x_i+b_i$ is
taken modulo 2.

\

\noindent The following technical lemma will be essential in the
derivation of our main result:

\

\begin {lemma} If $A,B$ are binary vectors of length $n$ and if $C$ is a
binary $[n,k]$ code, then
$$\Pi (C,A,B)= \langle \Psi _C | \sigma _{A,B} |\Psi _C \rangle
= \frac {e^{\frac {i\pi }{2} (A|B)}}{2^k}\sum _{X,Y\in C}e^{i\pi
(A|X)}\langle Y|X+B\rangle. \leqno (8)$$
\end {lemma}

\

\begin {proof} Indeed, (8) is an immediate consequence of (5) together
with (7) and the fact that $\Pi (C,A,B)= \langle \Psi _C | \sigma
_{A,B} |\Psi _C \rangle $.
\end{proof}

\

From (8), taking into account the fact tat the kets $|X\rangle $
form an orthonormal basis in the $2^n$-dimensional Hilbert space
$H^{\otimes n}$, it follows that if the binary vector $B$ is not a
codeword of $C$ then $\Pi (C,A,B) = 0$. Indeed, if $B\notin C$
then the linear code $C$ and the coset $C+B$ are disjoint, thus
the inner products $\langle Y|X+B\rangle $ appearing in (8) are
all zero.

\

If $B\in C$, then for every $X$ in (8) there exists exactly one
$Y$ such that the inner product $\langle Y|X+B\rangle $ is 1 (all
others being zero). Thus, the expectation value $\Pi (C,A,B)$ can
be expressed as follows:

$$\Pi(C,A,B)= \frac {e^{\frac {i\pi }{2} (A|B)}}{2^k}\sum _{X\in C}e^{i\pi
(A|X)}.\leqno(9)$$

\

\noindent Note that the sum appearing in (9) is zero if $A\notin
C^{\perp }$ and $2^k$ if $A\in C^{\perp }$ (for information on
linear codes and duality, see, for example, [5], Chapter
3).Therefore, if $A\notin C^{\perp }$ and $B\in C$ and then $\Pi
(C,A,B) = 0$, while if $A\in C^{\perp }$ and $B\in C$, then $\Pi
(C,A,B) = e^{\frac {i\pi }{2} (A|B)}$. Note that in the last case
the quantity $(A|B)$ will be an even integer, which makes the
quantity $e^{\frac {i\pi }{2} (A|B)}$ either 1 or $-1$.

\

By putting together the above facts, we get our main result.

\

\begin {theorem} If $A,B$ are binary vectors of length $n$ and if $C$ is
a binary $[n,k]$ code, then:

\

\item \ \ \ \ \ \ $\Pi (C,A,B)=0$ {\rm if} $B\notin C$,

\item \ \ \ \ \ \ $\Pi (C,A,B)=0$ {\rm if} $B\in C$ and $A\notin
C^{\perp}$,

\item \ \ \ \ \ \ $\Pi (C,A,B)=e^{\frac {i\pi }{2} (A|B)}$ {\rm
if} $B\in C$ and $A\in C^{\perp }$.

\end {theorem}

\

\section {Special  cases and nonlinear codes}

\

We will consider two special cases of the above Theorem 3.2.
First, consider the case $B=\mathbf{1}=(1,1,...,1)$. This
corresponds to measurements of the $x$- or $y$- spin components
for all the particles in the system described by $|\Psi _C\rangle
$. In this situation, Theorem 3.2 reduces to the following result
obtained in [1], detailing the expectation values
$\Pi(C,A):=\Pi(C,A,\mathbf{1})$.

\

\begin {theorem} If $A$ is a binary vector of length $n$ and if $C$ is
a binary $[n,k]$ code, then:

\

\item \ \ \ \ \ \ $\Pi (C,A)=0$ {\rm if} ${\mathbf{1}}\notin C$,

\item \ \ \ \ \ \ $\Pi (C,A)=0$ {\rm if} ${\mathbf{1}}\in C$ and
$A\notin C^{\perp}$,

\item \ \ \ \ \ \ $\Pi (C,A)=1$ {\rm if} ${\mathbf{1}}\in C$,
$A\in C^{\perp }$ {\rm and} $wt(A)\equiv 0 \ ({\rm mod} \
    4)$,

\item \ \ \ \ \ \ $\Pi (C,A)=-1$ {\rm if} ${\mathbf{1}}\in C$,
$A\in C^{\perp}$ {\rm and} $wt(A)\equiv 2 \ ({\rm mod} \ 4)$.
\end {theorem}

\

Note that in this care the "entanglement effect" $\Pi (C,A)\neq 0
$ appears precisely when $|\Psi _C\rangle $ has nonzero components
along both "$GHZ$ kets" $|00...0\rangle $ and $|11...1\rangle$.

\

The second special case we will address is
$B=\mathbf{0}=(0,0,...,0)$. This corresponds to measurements of
the $z$- spin components for the particles in a selected subset of
the set of entangled particles described by $|\Psi _C\rangle $. In
this case $B=\mathbf{0}\in C$ and $e^{\frac {i\pi }{2} (A|B)}=1$,
so that Theorem 3.2 reduces to the following result detailing the
expectation values $P(C,A):=P(C,A,\mathbf{0})$.

\

\begin {theorem} If $A$ is a binary vector of length $n$ and if $C$ is
a binary $[n,k]$ code, then $P (C,A)=0$ if $A\notin C^{\perp}$ and
$P (C,A)=1$ if $A\in C^{\perp }$.
\end {theorem}

\

The case of nonlinear binary codes can be approached in a similar
way. A binary nonlinear $(n,M)$-code is simply a $M$-element
subset $C\subset \{ 0,1\}^n$. The definition of the operators
$\sigma _{A,B}$ remains the same. The corresponding codeket will
be
$$|\Psi_C\rangle :=\frac {1}{\sqrt M}\sum_{(x_1,...,x_n)\in C}|x_1\rangle \otimes |x_2\rangle \otimes ... \otimes |x_n\rangle .$$
The analogue of Lemma 3.1 will be, in the nonlinear case, the
following proposition in which the expectation values $\Pi
(C,A,B)$ are expressed as incomplete exponential sums over sets of
the form $C\cap (C+B)$.

\

\begin {lemma} If $A,B$ are binary vectors of length $n$ and if $C$ is a
binary $(n,M)$-code, then

$$\Pi (C,A,B)= \frac {e^{\frac {i\pi }{2} (A|B)}}{M}\sum _{X\in C\cap (C+B)}e^{i\pi
(A|X)}. \leqno (10)$$
\end {lemma}

\begin{proof}

The relation $$\Pi (C,A,B)= \langle \Psi _C | \sigma _{A,B} |\Psi
_C \rangle = \frac {e^{\frac {i\pi }{2} (A|B)}}{M}\sum _{X,Y\in
C}e^{i\pi (A|X)}\langle Y|X+B\rangle. \leqno (11)$$ holds for
every binary $(n,M)$-code and every choice of binary vectors $A$
and $B$ and can be proved in exactly the same way as (8). To get
(10) note that for every $Y\in C$ the term $\langle Y|X+B\rangle$
appearing in (11) is 1 if $X=B-Y(=B+Y\ {\rm mod \ 2})\in C$ (case
in which $X\in C\cap (C+B)$ and 0 otherwise. Thus, (10) expresses
the expectation value $\Pi (C,A,B)$ as an incomplete exponential
sum over the set $C\cap (C+B)$.
\end {proof}

\

The following bound for the expectations $\Pi (C,A,B)$ is an
immediate consequence of Lemma 4.3.

\

\begin {corollary}
If $A,B$ are binary vectors of length $n$ and if $C\subset \{
0,1\}^n$ is a binary code, then $$|\Pi (C,A,B)|\leq \frac {|C\cap
(C+B)|}{|C|}.\leqno (12)$$
\end {corollary}

\

Note that if we use the inequality (12) in the special case that
$C\cap (C+B)=\emptyset $ we get the following result.

\

\begin {corollary}
If $A,B$ are binary vectors of length $n$ and if $C\subset \{
0,1\}^n$ is a binary code such that $C\cap (C+B)=\emptyset $, then
$$\Pi (C,A,B)=0.$$
\end {corollary}

\

For example, if we consider the above corollary with $C$ linear
and $B=(1,1,...,1)\notin C$ (which makes the coset $C+B=C+\mathbf
{1}$ disjoint from $C$) then we recover $\Pi(C,A)=\Pi
(C,A,\mathbf{1})=0$ (already proved in Theorem 4.1).

\

Finally, the following proposition presents an interesting
algebraic identity which is a consequence of (10).

\begin {proposition} Let $A$, $B$ be binary strings of length
$n$ such that $$(A|B)=1\ {\rm (mod 2)},$$ and let $C$ be a subset
of $\{ 0,1\}^n$. Then the following identity holds true:
$$\sum _{X\in C\cap (C+B)}e^{i\pi
(A|X)}=0.$$
\end{proposition}

\

\begin{proof}
If $(A|B)=1$ {\rm (mod 2)}, then $e^{\frac {i\pi }{2} (A|B)}=\pm
i$. On the other hand, $$\sum _{X\in C\cap (C+B)}e^{i\pi (A|X)}$$
is a real number. If $\sum _{X\in C\cap (C+B)}e^{i\pi (A|X)}\neq
0$ then the right hand side of (10) will be a purely imaginary
quantity, which is in contradiction to the fact that the
expectation value $\Pi (C,A,B)$ is a real number, indeed an
element of the interval $[-1,1]$. This concludes the proof.
\end{proof}

\

\

\noindent {\bf REFERENCES}

\

\

\noindent [1] M. Caragiu, {\it A Note on Codes and Kets}, Siberian
Electronic Mathematical Reports 2 (2005), 79-82.

\

\noindent [2] C. Cohen-Tannoudji, B. Diu and F. Laloe, {\it
Quantum Mechanics} (Volume I), Wiley-Interscience, 1996.

\

\noindent [3] D.M. Greenberger, M.A. Horne, A. Shimony and A.
Zeilinger, {\it Bell's theorem without inequalities, Amer. J.
Phys.} 58 (1990), 1131-1143.

\

\noindent [4] D.M. Greenberger, M. Horne and A. Zeilinger, in {\it
Bell's Theorem, Quantum Theory and Conceptions of the Universe},
M. Kafatos (editor), Kluwer Academic, Dordrecht, 1989

\

\noindent [5] J. H. van Lint, {\it Introduction to Coding Theory},
Third Edition, Springer Verlag, 1999.

\

\

\noindent Author's contact information:

\

\noindent Mihai Caragiu

{\it

\noindent 262 Meyer Hall

\noindent Department of Mathematics, Ohio Northern University

\noindent Ada, OH 45810

\noindent E-mail: m-caragiu1@onu.edu

}

\end {document}